# Invasion depth estimation of gastric cancer in early stage using circularly polarized light scattering: Phantom studies

M. R. Maskey[*a], M. Fukuno[a], A. Saito[a], A. Esumi[a], T. Kuchimaru[b] and N. Nishizawa[a]

[a]Department of Physics, School of Science, Kitazato University, 1-15-1 Kitazato, Sagamihara, Kanagawa, Japan 252-0373; [b]Center for Molecular Medicine, Jichi Medical University, 3311-1 Yakushiji, Shimotsuke-shi, Tochigi, Japan 329-0498

**ABSTRACT**

Depolarization of circularly polarized light due to multiple scattering in turbid media provides size distributions of scatterer. Applied it to the biological tissues as turbid media, scatters correspond to cell nuclei, which is abnormally grown in cancerous tissues. Therefore, the invasion depth of early-staged cancer can be estimated by comparisons of circular polarization of scattered light. In this study, we fabricated the optical phantoms made of resin and polystyrene beads to verify this technique by systematic experiments. The single-layered phantoms containing only one size of beads exhibits uniform monodispersed scattering media. Polarization images taken by a circular polarization imaging camera show systematically changes to the numerical density of scatters. Healthy and cancerous tissue phantoms exhibiting the lengths of mean free paths close to actual biological tissues were stacked to be bi-layered phantoms which imitates early-staged cancers. The averaged circular polarization values obtained from the images captured with the polarization camera are indicative of obvious changes depending on the thickness of cancerous layer of phantoms.

**Keywords:** Circularly polarized light, multiple scattering, optical biopsy, early-staged cancer

## 1. INTRODUCTION

Gastric cancers have been classified according to the invasion depth of tumor [1]. Early gastric cancers categorized in Tis and T1 stages can be removed without sacrificing the large part of organ by using endoscopic surgical procedures, while the more-advanced cancer have been treated by surgical treatment. Therefore, accurate tumor invasion depth in gastric cancer is of primary importance in deciding the therapeutic approach. Optical (endoscopic) techniques such as image-enhanced endoscopy (IEE) and endoscopic ultrasonography (EUS) has been used for detecting a tumor on the surface of stomach wall through emphasized abnormal blood vessels and mesh patterns [2-4]. Near-surface information obtaining these techniques can be combined with invasion depth via machine learning, which is progressing recently [5,6]. In contrast, a direct evaluation method of invasion depth is required.

We proposed optical, direct method for invasion depth evaluation using scattering circularly polarized light (CPL). When circularly polarized light beams impinge on a biological tissue, they penetrate and propagate into the tissue and are scattered multiple times by cell nuclei which are the main scatterers in tissues. Depolarization of the resultant scattered light provides valuable information regarding cell nuclei, because the depolarization strength strongly depends on the size of the nucleus [7-9]. This method, called "Circularly polarized light scattering (CPLS)" method, can be applied on cancer diagnosis by detected enlarged cell nuclei due to canceration [10, 11]. In CPLS method with infrared CPL, the scattering volume of biological tissue reaches to from 2 to 3 mm. The resultant circular polarization of scattered light depends on the ratio of the cancerous layer volume to the entire scattering volume. To date, we have reported computational analysis of the quantitative estimation of cancer invasion depth using CPLS method [12]. Also, we have prepared artificially bi-layered cancerous/ normal tissues [13]. However, biological tissues include many kinds of scatterers other than cell nuclei. They may cause the depolarization of circularly polarized light. In this paper, we fabricated ideal biological phantoms for CPLS experiments to verify the depolarization dependence of cancer layer thickness.

## 2. EXPERIMENTAL METHOD

### 2.1 Phantom preparation

We fabricated biological phantoms by ultraviolet (UV) irradiation of UV curing resin (3094F；ThreeBond Co., Ltd.) with polystyrene beads. The refractive indices of polystyrene (PS) and resin is $n_{PS} = 1.58$ and $n_{resin} = 1.55$. The relative refractive index $m = n_{PS}/n_{resin} = 1.02$, which is nearly equivalent to the relative refractive index of cell nucleus ($n_n$) and cytoplasm ($n_c$) [14], $m = n_n/n_c = 1.04$. The PS beads of 5.0 μm diameter (SX-500H；Soken Chemical & Engineering Co., Ltd.) and of 12.0 μm diameter (SBX-12 ; Sekisui Kasei Co., Ltd.) were used for imitating healthy and cancerous tissues, respectively. These sizes correspond to the typical size of a nucleus [10, 15] and average values experimentally measured in biological specimens [11].

The numerical density $N$ of a number of PS beads for resin volume is varied. The mean free path ($L$) of light in biological tissues can be estimated by the optical coefficients, absorption and scattering coefficients ($\mu_a$ and $\mu_s$), yielding $L = 97.3$ μm for 950 nm. Under the assumption that scattering particle is immovable and regards as a point, the $L$ value can be converted to $N$ by $N = 1/L\sigma$ where $\sigma$ is a scattering cross section. When $L = 97.3$ μm, we obtained $N = 5.20 \times 10^8$/mL for PS beads of 5.0 μm diameter. On the other hand, considering the cross section of particle, the definition is $N = 1/4\sqrt{2}L\sigma$, obtaining $N = 0.91 \times 10^8$/mL. In this study, the $N$ values are varied between these values, $5.2, 2.0, 1.0$ and $0.91 \times 10^8$/mL. Given that the intercellular distance between cell nuclei does not change a lot by canceration, the $L$ values in healthy and cancerous tissue phantoms were almost the same. For cancerous tissue phantoms, the $N$ values are varied $0.90, 0.35, 0.17$ and $0.16 \times 10^8$/mL. Table 1 shows the list of fabricated single-layered phantoms using the $L$ values converted by $L = 1/N\sigma$ as the common parameters for heathy and cancerous phantoms.

Table 1. Sample list of single-layered phantoms

| $N$ [$\times 10^8$/mL] | mean free path ($L$) [μm] ($L = 1/N\sigma$) | | | |
|---|---|---|---|---|
| | 97 | 255 | 509 | 554 |
| | no beads(∞) | | | |
| Healthy ($2a = 5$ μm) | 5.20 | 2.00 | 1.00 | 0.91 |
| Cancer ($2a \approx 12$ μm) | 0.90 | 0.35 | 0.17 | 0.16 |

Phantoms for invasion depth estimation have two-layer structure which is composed of a cancer layer with $t$ of thickness and a healthy layer. The cancerous tissue phantoms thinned to approximately $t, t = 0, 0.5, 1.0, 1.5$ and $3.0$ mm, by mechanical polishing were stuck on the healthy tissue phantoms, and both layers were adhered with resin. Total thicknesses of the phantoms were fixed to be 3.0 mm or more. Subsequently, the surface of the two-layer phantoms were mechanically planarized.

## 2.2 Optical coefficients

The optical coefficients of the single-layered phantoms, absorption and scattering coefficients ($\mu_a$ and $\mu_s$), were evaluated by the reverse Monte Carlo method (reverse MC) [16, 17]. Absorption and reflection spectra obtained with the integrating sphere with a range from 600 to 1000 nm. A plausible value of the optical coefficient was assumed and introduced into a conventional Monte Carlo algorithm for optical scattering to obtain absorption and reflection spectra. Assigning optical coefficient varied slightly to make the calculation result asymptotic to the experimental values. The optical coefficients of actual biological tissues were obtained using semiempirical formulae [12, 18, 19] and experimental measurements obtained from the stomach wall and its carcinoma [20, 21]. The optical coefficients obtained through reverse MC calculations are compared with the actual values, and feedback was applied to the $N$ values.

## 2.3 Monte Carlo simulation for polarized light scattering

To investigate the polarization and passage distribution of scattered light, the polarization-light MC algorithm developed by Rammella-Raman *et al.* [22] was adopted in this study. The polarization state of light is expressed using the Stokes vector $\boldsymbol{S}$, $\boldsymbol{S} = (S_0, S_1, S_2, S_3)^T$, where $S_0, S_1, S_2$ and $S_3$ are the Stokes polarization parameters [23], and the degree of circular polarization (DOCP) is defined as $S_3/S_0$. Polarized MC simulations were carried out for pseudo-biological tissues with a bilayered structure: a cancerous layer on a healthy layer. The pseudo-cancerous and healthy tissues comprise of an aqueous dispersion of particles with diameters 5.0 and 12.0 μm, respectively. The cancerous layer with various thicknesses $t$ is located on a healthy layer, imitating cancer progression from the surface to the interior of the tissues. A schematic configuration of the CPL source, detector, and pseudo-tissues is illustrated in Figure 1(a). The wavelengths of incident CPL were chosen to be 600 and 950 nm. Both wavelengths showed significant differences in the DOCP of light scattered from the cancerous and healthy tissues, coupled with the large intensities. The incident angle was fixed at 1°, and the detection angle ϕ was varied from 30° to 50°. The CPL detector collected the scattered outgoing light beams from pseudo-tissues and over a detection area of width = 1 mm located 1 mm from the point of incidence. The number of trials (incident photons number) for this calculation was 800,000.

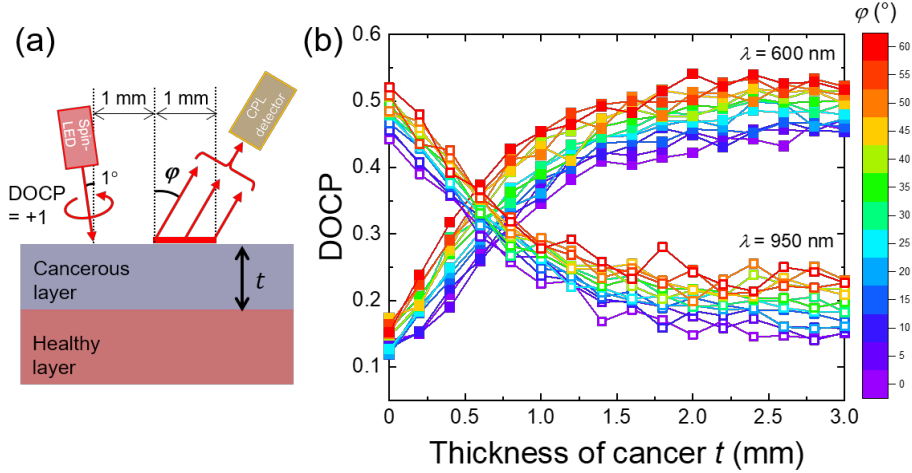

Figure 1: (a) Schematic of optical geometry and layered structure of the pseudo-tissues in which a cancerous layer lying on the surface progresses deeper. (b) The calculated resultant DOCP values of light scattered from pseudo-tissues as a function of thickness of cancer $t$ with different detection angles $\varphi$ for $\lambda = 950$ nm (opened squares) and $\lambda = 600$ nm (closed squares). (Adapted from [12])

## 2.4 Optical experiments

The DOCP values of the scattered light were experimentally measured. Incident light with wavelengths of 617 and 850 nm was obtained using LED with 1.0 and 1.6 W (M617L5 and M850LP1; Thorlabs, Inc.), respectively and converted to right-handed CPL using a linear polarizer and quarter-wave plate (QWP). Light scattered from the sample was converted to linear polarized light with a QWP and detected using a polarization imaging camera (Polarsens; Sony Semiconductor Solutions Corp.). The polarization imaging camera having 2 × 2 microgrid polarimetric array pattern enables snapshot polarimetric imaging by spatially multiplexing the wire grid analyzer oriented in 0°, 45°, 90°, and 135° degree orientations. The Stokes vector $\boldsymbol{S'}$ of light passing through a QWP whose fast axis is at an angle $\theta$ is calculated by Müeller matrix $\boldsymbol{M}_{\mathrm{QWP}}(\theta)$ [23];

$$\boldsymbol{S'} = \boldsymbol{M}_{\mathrm{QWP}}(\theta) \cdot \boldsymbol{S}$$

$$\begin{pmatrix} S_0' \\ S_1' \\ S_2' \\ S_3' \end{pmatrix} = \begin{pmatrix} 1 & 0 & 0 & 0 \\ 0 & \cos^2 2\theta & \sin 2\theta \cos 2\theta & \sin 2\theta \\ 0 & \sin 2\theta \cos 2\theta & \sin^2 2\theta & -\cos 2\theta \\ 0 & -\sin 2\theta & \cos 2\theta & 0 \end{pmatrix} \begin{pmatrix} S_0 \\ S_1 \\ S_2 \\ S_3 \end{pmatrix}. \quad (1)$$

When the fast axis of a QWP is in the vertical axis, $\theta = \pi/2$,

$$\begin{pmatrix} S_0' \\ S_1' \\ S_2' \\ S_3' \end{pmatrix} = \begin{pmatrix} +1 & 0 & 0 & 0 \\ 0 & +1 & 0 & 0 \\ 0 & 0 & 0 & +1 \\ 0 & 0 & -1 & 0 \end{pmatrix} \begin{pmatrix} S_0 \\ S_1 \\ S_2 \\ S_3 \end{pmatrix} = \begin{pmatrix} +S_0 \\ +S_1 \\ +S_3 \\ -S_2 \end{pmatrix}. \quad (2)$$

Therefore, the circular polarization component, $S_3$, can be obtained as the differences in the photon intensities detected at a pixel-sized analyzer with 45° and 135° orientation, $I_{45°} - I_{135°}$, whereas the preponderance of the horizontal linear polarization over the vertical linear polarization, $S_1$, can be detected as $I_{0°} - I_{90°}$. In this study, the image data obtained with the polarization camera was divided to four polarized images according to four orientations, then $I_{45°} - I_{135°}$, $I_{45°} + I_{135°}$, $I_{0°} + I_{90°}$ images are obtained, and finally, DOCP images obtained as

$$\frac{2(I_{45°} - I_{135°})}{(I_{45°} + I_{135°}) + (I_{0°} + I_{90°})}. \quad (3)$$

This DOCP calculation is based on an assumption that a whole detected light is polarized and is incorrect for partially polarized light which includes unpolarized light component. An accurate DOCP values according to its definition $S_3/S_0$ can be calculated by multiplying the obtained DOCP values by the degree of polarization $P$. However, $P$ value should be separately measured. Therefore, we use the equation (3) in this study, for we put an importance not on the mathematical accuracy but on the comparison of its magnitude. The incident angle ($\theta$) and detection angle ($\varphi$) are defined as shown Figure 2, and the polarization images were obtained at $(\theta, \varphi) = (0°, 30°)$ and $(30°, 0°)$. According to the calculation results, the 600 and 950 nm in wavelength should be adopted [12]. However, the polarization camera used in this study has low sensitivity at longer than 900 nm in wavelength. Therefore, instead of 950 nm, we employed the incident CPL of 850 nm which is also expected to exhibit a significant difference in the DOCP between cancerous and healthy tissues.

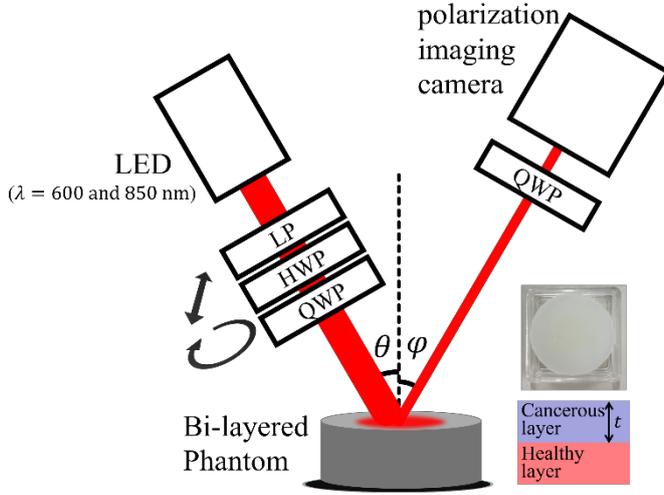

Figure 2: A schematic illustration of experimental setup for taking the images of the degree of circular polarization (DOCP) in this study. The incident CPL emitted from LED through a linear polarizer (LP), a half-wave plate (HWP) and a quarter wave plate (QWP), irradiates to the phantoms located at the origin. The incident and detection angles, $\theta$ and $\varphi$ are defined as angles form the normal line on the origin. The right upper inset is a photograph from the top and the lower inset represents schematically the two-layered phantom structure.

For comparison, the circular polarization images were captured with a polarization-probe polarization-imaging system (V3PO, OPT Gate Co., Ltd.) [24]. This system has a liquid crystal polarization diffraction grating in front of an image sensor to spatially separate the light into its right- and left-handed CPL components. The distributions of DOCP values can be obtained as differences between them. The optical angular configuration was $(\theta, \varphi) = (0°, 30°)$ and the wavelength of the incident CPL is 940 nm.

## 3. RESULTS

### 3.1 Monte Carlo simulations [12]

Figure 1(b) shows the calculated DOCP values of light scattered from the pseudo-tissues as a function of thickness of cancer $t$ with different detection angles $\varphi$ for $\lambda = 600$ nm (closed squares) and $\lambda = 950$ nm (open squares). The relationship of DOCP with $t$ showed opposing tendencies between 600 and 950 nm wavelengths. When $t$ increased from 0 to approximately 1.0 mm, the DOCP values increased and decreased monotonically for 600 and 950 nm, respectively. In this thickness region, most of the light beams dive under the underlaid healthy layers. Therefore, the DOCP values varied depending on the ratio of cancerous layer volume to the entire sampling volume. When $t$ increased further, the saturation behavior varied depending on the detection angle $\varphi$. The DOCP values for large $\varphi$ saturated at 1.4 mm, while the values for small $\varphi$ continued to vary till 2.0 mm. The sampling depth depends on the detection angle $\varphi$. When $\varphi$ was small (nearly perpendicular), the detected light contained several light beams that dived deeper. Conversely, when $\varphi$ was large, most of the light beams scattered over a shallow volume. Therefore, the difference in DOCP values between $\varphi$ was derived from the observation that the scattering volume of light with small $\varphi$ included healthy tissues, whereas the scattering volume of light with large $\varphi$ was fulfilled by the cancerous layer. Further increases in $t$ resulted in the saturation of the DOCP values over the whole angular range since the scattering volumes were fulfilled by the cancer layer. In summary, the depth profile can be obtained from the differences in DOCP values for wavelengths $\lambda$ and detection angles $\varphi$ for a cancer layer thinner and thicker than 1.0 mm, respectively. As a result, the deepest depth for quantitative cancer thickness measurement was approximately 2.0 mm.

## 3.2 Single-layered phantoms

The microscope observations of the fabricated phantoms show mostly monodispersed PS beads in the resin matrix. For the phantoms with small $N$, the mean free paths obtained from the reverse MC method and optical experiments with an integrating sphere are half or one-thirds shorter than the estimated values from the numerical densities. The flocculated PS beads gives the longer $L$ values, which cannot cause the smaller $L$.

Figure 3(a) shows the DOCP images measured with a polarization imaging camera with an irradiation of 617-nm CPL with $(\theta, \varphi) = (30°, 0°)$. The images indicate the DOCP distributions for the inadequate flattening process. Therefore, the DOCP values of the flat part were extracted and averaged. Figure 3(b) shows the averaged DOCP values as a function of $L$. the DOCP values are monotonically increased with an increase of $L$, which means the decrease of the number of scattering events. The scattered light from the phantoms with the smallest numerical densities ($L = 554$ µm) is drastically decreased, resulting in the inconsistent DOCP values with the monotonical increasing behavior. The DOCP values from healthy tissue phantoms are larger than that from cancerous tissue phantoms, which are consistent with the MC simulations for polarized light scattering [12] and derived from difference in particle sizes. These results indicate that the definition excluding the cross section of particle is appropriate for the phantoms. Hereafter the $N$ values of $5.20$ and $0.90 \times 10^8$/mL were adopted for healthy and cancerous tissues phantoms, respectively.

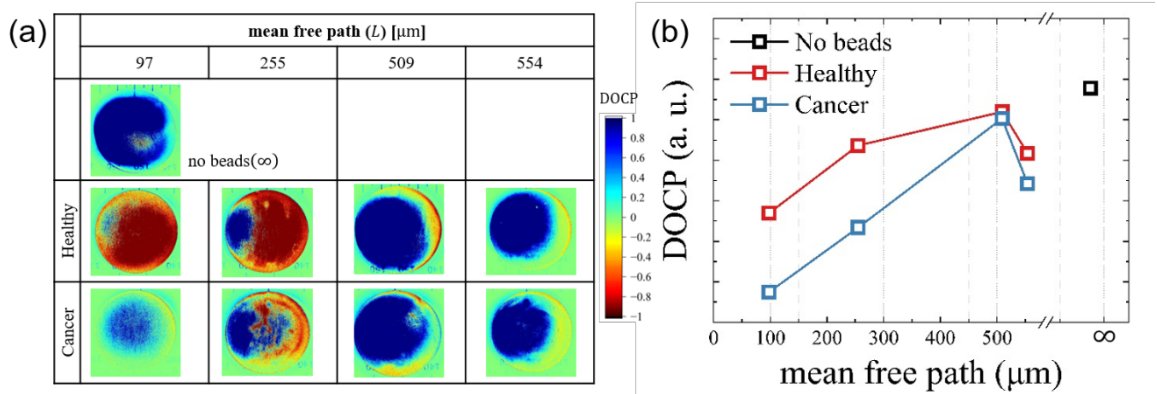

Figure 3: (a) Circular polarization images of the single-layered phantoms obtained with a polarization imaging camera. Together with a resin without beads (the upper row), the DOCP images of the phantoms imitating healthy tissue (middle row) and cancerous tissue (lower row) phantoms. The right-handed CPL (DOCP = +1) of 617 nm in wavelength irradiated with 30° of incident angle. (b) The mean free path length dependence of the DOCP values of light scattered from phantoms without a bead (black plots and lines), with beads imitating healthy (red) and cancerous (blue) tissue phantoms.

## 3.3 Bi-layered Phantoms

Figure 4 shows the experimental results for bi-layered phantoms. Fig. 4(a) shows the DOCP images of phantoms taken by the Polarsens system (upper row), together with the DOCP images of artificial bi-layered tissues taken by the V3PO system (lower row) for comparison. The averaged DOCP values obtained from the center parts of these images as a function of the thickness of cancer layer, $t$, are shown in Fig. 4(b). The red and blue points and lines denotes the values obtained at $(\theta, \varphi) = (0°, 30°)$ and $(30°, 0°)$, respectively. No marked difference is apparent between them. The DOCP values are notably changed between $t = 1.0$ and $1.5$ mm, which are consistent with the V3PO observation results for the biological tissues [25]. In the previous study [8], the MC simulations indicates that 80-% photons dive into 1.3~1.5 mm deep and 55-% photons penetrate into a depth of 2.0 mm in case that light beams enter vertically and then outgoing from the surface with an angle of 30°. Therefore, for the phantoms having a cancer layer thinner than 1.0 mm, most of photons reach the under healthy layer, resulting in comparatively high DOCP values, whereas, for the phantoms having the thicker cancerous layer, the lower DOCP values are shown due to the photons which undergo scattering events with larger particles. However, such differences are a little bit small. In the experimental setup, the incident angle of light is not completely collimated and the cameras can detect the light with perpendicular within the angular region $\pm 3°$. Therefore, the experimental results show a somewhat ambiguity derived from these angle distributions. We would like to make these the next challenges.

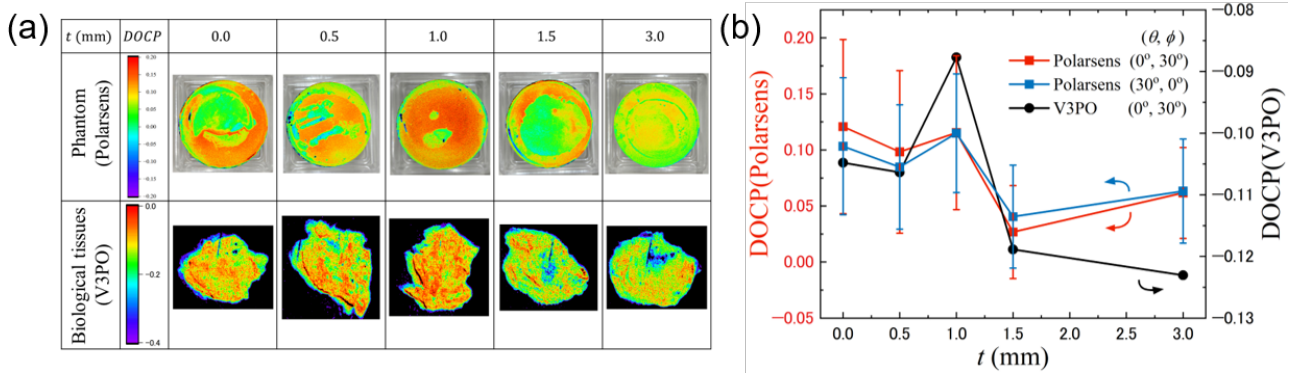

Figure 4: (a) Circular polarization images of the bi-layered phantoms obtained with a polarization imaging camera (upper row), together with the images of artificial bi-layered tissues taken by the other polarization camera, V3PO [25] (rower row). The right-handed CPL (DOCP = +1) of 617 nm in wavelength irradiated with 30° of incident angle. (b) The $t$ dependence of the DOCP values of bi-layered phantoms measured with the Polarsens system at $(\theta, \varphi) = (0°, 30°)$ (red squares and lines) and $(30°, 0°)$ (blue) on the left axis. For comparison, the DOCP values of biological tissues taken by V3PO system are also plots by black circles and lines on the right axis.

## 4. CONCLUSIONS

We investigated the applicability of CPLS method for the invasion depth estimation. We fabricated the optical phantoms made of UV curing resin and polystyrene beads imitating cytoplasm and cell nuclei, respectively. The single-layered phantoms containing only one size of beads exhibits uniform monodispersed scattering media. They shows DOCP values according to numerical density of containing particles. Length of the optical mean free path in actual tissues are approximately 97 μm. To realize it in the phantoms, the $N$ values of 5.20 and $0.90 \times 10^8$/mL are appropriate for healthy and cancerous tissues phantoms, respectively. Using these values, we fabricated the two-layered phantoms imitating cancer progressing from a surface to deeper. The distributions of DOCP values are captured with circular polarization imaging camera. The averaged DOCP values obtained from the captured images show the notably changes between 1.0 and 1.5 mm of the thickness of cancerous layer, which is also observed in the actual two-layered biological tissues with the other polarization camera. These results indicate that there is room for further improvements: averaging process from polarization images to DOCP values, flattening and bonding process of phantoms, and optical alignment of polarization camera.

## REFERENCES

[1] Wu, H.-Q., Wang, H.-Y., Xie, W.-M., Wu, S.-L., Li, Z.-F., Zhang X.-M., and Li H., "Scanning photoacoustic imaging of submucosal gastric tumor based on a long focused transducer in phantom and in vitro experiments," J. Innov. Opt. Health Sci. 12, 1950011 (2019).
[2] Emura, F., Saito, Y., and Ikematsu, H., "Narrow-band imaging optical chromocolonoscopy: Advantages and limitations," World J. Gastroenterol. 14, 4867 (2008).
[3] Kikuchi, D., Iizuka, T., Hoteya, S., Yamada, A., Furuhata, T., Yamashita, S., Domon, K., Nakamura, M., Matsui, A., and Mitani, T., "Prospective Study about the Utility of Endoscopic Ultrasound for Predicting the Safety of Endoscopic Submucosal Dissection in Early Gastric Cancer (T-HOPE 0801)," Gastroenterol. Res. Pract. 2013, 329385 (2013).
[4] Yagi, K., Nakamura, A., Sekine, A., and Umezu, H., " Magnifying endoscopy with narrow band imaging for early differentiated gastric adenocarcinoma," Dig. Endosc., 20, 115 (2008).
[5] Nagao, S., Tsuji, Y., Sakaguchi, Y., Takahashi, Y., Minatsuki, C., Niimi, K., Yamashita, H., Yamamichi, N., Seto, Y., Tada, T., and Koike, K., "Highly accurate artificial intelligence systems to predict the invasion depth of gastric cancer: efficacy of conventional white-light imaging, nonmagnifying narrow-band imaging, and indigo-carmine dye contrast imaging," Gastrointest. Endosc, 92, 866 (2020).

[6] Zhu, Y., Wang, Q.-C., Xu, M.-D., Zhang, Z., Cheng, J., Zhong, Y.-S., Zhang, Y.-Q., Chen, W.-F., Yao, L.-Q., Zhou, P.-H., Li, and Q.-L., "Application of convolutional neural network in the diagnosis of the invasion depth of gastric cancer based on conventional endoscopy" Gastrointest. Endosc, 89, 806 (2019).

[7] Bickel, W. S., Davidson, J. F., Huffman, D. R., and Kilkson, R., "Application of polarization effects in light scattering: A new biophysical tool," Proc. Nat. Acad. Sci USA 73, 486 (1976).

[8] Nishizawa, N., Hamada, A., Takahashi, K., Kuchimaru. T., and Munekata, H., "Monte Carlo simulation of scattered circularly polarized light in biological tissues for detection technique of abnormal tissues using spin-polarized light emitting diodes", Jpn. J. Appl. Phys. 59, SEEA05 (2020).

[9] Nishizawa, N., Esumi, A., and Ganko, Y., "Depolarization diagrams for circularly polarized light scattering for biological particle monitoring", J. Biomed. Opt. 29, 075001 (2024).

[10] Kunnen, B., Macdonald, C., Doronin, A., Jacques, S., Eccles, M., and Meglinski, I., "Application of circularly polarized light for non-invasive diagnosis of cancerous tissues and turbid tissue-like scattering media," J. Biophotonics 8, 3317 (2015).

[11] Nishizawa, N., Al-Qadi, B., and Kuchimaru, T., "Angular optimization for cancer identification with circularly polarized light," J. Biophotonics 14, 202000380 (2020).

[12] Nishizawa, N., and Kuchimaru, T., "Depth estimation of tumor invasion in early gastric cancer using scattering of circularly polarized light: Monte Carlo Simulation study", J. Biophotonics 15, 202200062 (2022).

[13] Nishizawa, N., and Kuchimaru, T., "Experimental depth estimation of cancer invasion via circularly polarized light scattering," Proc. . Biophotonics 15, 202200062 (2022).

[14] Liu, H., Beauvoit, B., Kimura, M., and Chance, B., "Dependence of tissue optical properties on solute-induced changes in refractive index and osmolarity" J. Biomed. Opt 1, 200 (1996).

[15] Backman, V., Gurjar, R., Badizadegan, K., Itzkan, I., Dasari, R. R., Perelman, L. T., and Feld, M. S., "Polarized Light Scattering Spectroscopy for Quantitative Measurement of Epithelial Cellular Structures *In Situ*," IEEE J. Sel. Top. Quantum Electron. 5, 1019 (1999).

[16] Wilson, B. C., Patterson, M. S., and Flock, S. T., "Indirect versus direct techniques for the measurement of the optical properties of tissues," Photochem. Photobiol. 46, 301 (1987).

[17] Moffitt, T., Chen, Y.-C., and Prahl, S. A., "Preparation and characterization of polyurethane optical phantoms," J. Biomed. Opt. 11, 041103 (2006).

[18] Alexandrakis, G., Rannou, F. R., and Chatziioannou, A. F., "Tomographic bioluminescence imaging by use of a combined optical-PET (OPET) system: a computer simulation feasibility study," Phys. Med. Biol. 50, 4225 (2005).

[19] Jacques, S. L., "Optical properties of biological tissues: a review," Phys. Med. Biol., 58, R37 (2013).

[20] Thueler, P., Charvet, I., Bevilacqua, F., Ghislain, M. S., Ory, G., Marquet, P., Meda, P., Vermeulen, B., and Depeursinge, C., "*In vivo* endoscopic tissue diagnostics based on spectroscopic absorption, scattering, and phase function properties," J. Biomed. Opt., 8, 495 (2003).

[21] S. A. Prahl. Oregon Medical Laser Clinic. http://omlc.ogi.edu/spectra/index.html

[22] Ramella-Roman, J. C., Prahl, S. A., Jacques, S. L., "Three Monte Carlo programs of polarized light transport into scattering media: part I,"Opt. Express 13, 4420 (2005).

[23] Collett, E., " Field Guide to Polarization," SPIE, Bellingham, WA (2005).

[24] Sakamoto, M., Nhan, H. T., Noda, K., Sasaki, T., Tanaka, M., Kawatsuki, N., and Ono, H., "Polarization-probe polarization-imaging system in near-infrared regime using a polarization grating" Sci. Rep. 12, 15268 (2022).

[25] Nishizawa, N., Maskey, M. R., Kuchimaru, T., Kobari, T., Kaneko, Y., Suzuki, M., Sakamoto, M., Tanaka, M., Ono, H., "Depth estimation of cancer invasion with polarization probe $S_3$ imaging using scattering of circularly polarized light," The JSAP spring meeting Paper 03-387, 24a-P03-4 (2024).